\documentclass[11pt]{article}
\usepackage{cite}
\usepackage{a4wide}
\usepackage{graphicx}
\usepackage{amsfonts}
\usepackage{amstext}
\usepackage{amsmath}
\usepackage{amssymb}
\usepackage{amsthm}
\usepackage[mathscr]{eucal}
\newtheorem{theorem}{Theorem}
\newtheorem{proposition}[theorem]{Proposition}

\newtheorem{corollary}[theorem]{Corollary}
\makeatletter
 
 \@addtoreset{equation}{section}
\makeatother
\def\a{\alpha}

\def\ba{{\boldsymbol\alpha}}
\def\bl{{\boldsymbol\ell}}
\def\C{\mathbb{C}}

\def\e{\epsilon}

\def\i{\infty}

\def\l{\lambda}

\def\N{\mathbb{N}}

\def\1{\bf{1}}
\def\R{\mathbb{R}}

\def\T{\mathbb{T}}
\def\th{\theta}

\def\Z{\mathbb{Z}}

\begin{document}
\title{On the $q$-TASEP with a random initial condition}

\author{Takashi Imamura
\footnote {Department of Mathematics and Informatics, 
Chiba University,~E-mail:imamura@math.s.chiba-u.ac.jp}
, Tomohiro Sasamoto
\footnote { Department of Physics,
Tokyo Institute of Technology,~E-mail: sasamoto@phys.titech.ac.jp}}

\maketitle

\begin{abstract}
When studying fluctuations of models in the 1D KPZ class including the ASEP and the $q$-TASEP, 
a standard approach has been to first write down a formula for $q$-deformed moments and 
constitute their generating function. This works well for the step initial condition, but 
there is a difficulty for 
a random initial condition (including the stationary case): 
in this case only the first few moments are finite and the rest diverge. 
In a previous work \cite{IS2017p},  we presented a method dealing 
directly with the $q$-deformed Laplace transform of an observable, in which 
the above difficulty does not appear. There the Ramanujan's 
summation formula and the Cauchy determinant for the theta functions
play an important role. In this note, we give an alternative approach 
for the $q$-TASEP without using them. 

\end{abstract}

\section{Introduction and results}
Recently remarkable progress has been made on interacting particle processes in the KPZ universality class
\cite{Corwin2012, QS2015, Sa2016}.  
In spite of their interaction of nonlinear nature, the limiting distributions and correlations of certain quantities have been 
identified for many models. Among them is the $q$-totally asymmetric simple exclusion process ($q$-TASEP), 
in which the $i$th particle hops to the right neighboring site on $\mathbb{Z}$ with the rate $a_i(1-q^{\rm gap})$
where $0\le q<1$ and $0<a_i\leq 1$ and the gap means the distance to the particle ahead. More precisely the rules of 
the $q$-TASEP for the case with $N$ particles are given as follows. 
Let $x_i(t),~i=1,2,\cdots,N$ be the position of the particle labeled $i$ at time $t\in \R_{\ge 0}$. A double occupancy 
at a site is prohibited and we put the order $x_1(t)>x_2(t)>\cdots >x_N(t)$.
Each particle hops to the right neighboring site with the rate of the $i$-th particle given by 
\begin{align}
a_i(1-q^{x_{i-1}(t)-x_i(t)-1}),
\label{11}
\end{align}
$1\leq i\leq N$, with $x_{0}(t)=\infty$ by convention. 
Note that $x_{i-1}(t)-x_i(t)-1$ means the gap (of empty sites) between $i$-th and the former $(i-1)$-th particle
and that the hopping rate (\ref{11}) depends on a configuration of the particles. 
As the gap becomes large, so does the rate and it approaches $a_i$ when the gap tends to infinity. 
On the other hand, if the two particles adjoin, the rate becomes zero, representing the exclusion interaction
among particles.
For the special case of $q=0$, the rate~\eqref{11} is always $a_i$ and does not depend on the gap. 
This is nothing but the TASEP (with particle dependent hopping rate $a_i$). 
The $q$-TASEP was introduced in~\cite{BC2014} as a marginal 
dynamics associated with the $q$-Whittaker process (though the dynamics of the gaps is called the $q$-boson 
totally asymmetric zero range process and had been introduced earlier in \cite{SW1998}) and has become 
one of the standard models for studying KPZ universality. Various generalized models have been introduced 
and studied since then,  see for instance \cite{Po2013, CP2015,BP2016p}. 

We are interested in the fluctuation properties of $x_N(t)$, the position of the $N$th particle. 
For the TASEP, with $q=0$, the limiting distribution of $x_N(t)$ has been identified 
for various initial conditions. First, for the step initial condition, which is given by  
\begin{align}
x_i(0)=-i,~i=1,2,\cdots,N,
\label{step}
\end{align}
it was shown in ~\cite{Johansson2000} that the limiting distribution is given by 
the GUE Tracy-Widom distribution~\cite{TW1994}. 
Then various other cases have been studied such as a random initial condition 
\cite{BR2000, IS2004, FS2006}, periodic initial condition~\cite{Sasamoto2005,BFPS2007}, 
and more general fixed initial condition \cite{MQR2017p, FO2017p}. 
All these results for TASEP (and for its cousin models like the polynuclear growth model)
have been obtained by using its connection to determinantal point processes,
which are related to the random matrix theory~\cite{Mehta2004,Forrester2010} and 
provide a unified approach. 

For general $q$, the same method does not work because a direct connection to a determiantal 
point process has not been found (see, however,  a few works in this direction 
\cite{IS2016, Borodin2016p,BO2017}). 
But a few methods have been devised and successfully been applied to the $q$-TASEP and its generalizations 
with many results. 
A standard approach for general $q$ has been to first write down a formula for $q$-deformed moments 
and constitute their generating functions.
For example, for the $q$-TASEP with the step initial condition (\ref{step}), 
the $n$-th $q$-deformed moment can be written as a multiple integral as 
\begin{equation}
\langle q^{n(x_N(t)+N)} \rangle 
=
\frac{(-1)^nq^{\frac12 n(n-1)}}{(2\pi i)^n}\int \prod_{1\leq j<k\leq n} \frac{z_j-z_k}{z_j-qz_k} \prod_{j=1}^n 
\left(\prod_{m=1}^N\frac{a_m}{a_m-z_j}\right) e^{(q-1)tz_j}\frac{dz_j}{z_j},
\label{mom_step}
\end{equation} 
where the contour for $z_j$ contains $\{ qz_k\}_{k>j}$ and $a_m$'s but not 0. 
This type of formula can be obtained by either using Macdonald operators \cite{BC2014} or 
duality \cite{BCS2014}. Next we compose a generating function of these moments. By using 
the $q$-binomial theorem, we find 
\begin{align}
\sum_{n=0}\frac{\zeta^n}{(q;q)_n}
\left\langle
q^{n(x_N(t)+N)}
\right\rangle
=
\left\langle \frac{1}{(\zeta q^{x_N(t)+N};q)_{\infty}}\right\rangle .
\label{qLapgen}
\end{align}
The right hand side is nothing but the $q$-Laplace transform of the pdf of $x_N(t)+N$,
which should contain the full information about the statistics of $x_N(t)$. 
In fact it turns out that the $q$-Laplace transform can be written as a Fredholm determinant 
and one can establish the Tracy-Widom law from this formula \cite{BC2014, BCS2014}. 

However, most results, obtained with this approach, have been restricted for the step 
initial condition (\ref{step}) due to a few technical difficulties.  
Let us consider, as a generalization of the step initial condition (\ref{step}), 
the following random initial condition, 
\begin{align}
-1-x_1(0)=X_1,~x_{i-1}(0)-x_{i}(0)-1=X_i \text{~for}~i=2,\cdots,N,
\label{rand}
\end{align}
where $X_1,\cdots,X_N$ are independent $q$-Poisson random variables
defined for $\a\in [0,a_k)$ by
\begin{align}
\text{Prob}(X_k=n)=(\a/a_k;q)_{\infty}\frac{(\a/a_k)^n}{(q;q)_n},
\label{16}
\end{align}
where $k\in (1,2,\cdots,N)$ and $n\in\Z_{\ge0}$. Note that the initial 
condition~\eqref{rand} means that the gap between $i$-th and $(i-1)$-th particle
is distributed as the $q$-Poisson random variable with parameter $\a/a_i$~\eqref{16} except for $i=1$,
for which case $X_1$ describes the gap between the particle and the origin. 
Note that, when $\a=0$, the rhs  of~\eqref{16} becomes $\delta_{n,0}$, then~\eqref{rand} reduces to the 
step initial condition~\eqref{step}.  
It is known that, for the $q$-TASEP with infinite number of particles and with $a_i=1$, 
if the gaps between the consecutive particles are given by the independent $q$-Poisson random variables with $\a$,
it is stationary. Because of this reason we call the initial condition (\ref{rand})  
the half stationary initial condition. By setting $a_i=1,i\geq 2$ and considering $a_1\to \alpha$ limit, 
one can study the stationary $q$-TASEP based on the analysis for the half-stationary initial condition
\cite{IS2017p}. 

In~\cite{BCS2014}, a nested contour integral representation of the
$n$-th moment  for this random initial condition was given, 
which is reproduced here, 
\begin{equation}
\langle q^{n(x_N(t)+N)} \rangle 
=
\frac{(-1)^nq^{\frac12 n(n-1)}}{(2\pi i)^n}\int \prod_{1\leq j<k\leq n} \frac{z_j-z_k}{z_j-qz_k} \prod_{j=1}^n 
\left(\prod_{m=1}^N\frac{a_m}{a_m-z_j}\right) e^{(q-1)tz_j}\frac{dz_j}{z_j-\a/q},
\label{mom_rand}
\end{equation} 
where the contour for $z_j$ contains $\{ qz_k\}_{k>j}$ and $a_m$'s but not $\a/q$. 
(Note, for this initial data, that  $\langle\cdot\rangle$ indicates the expectation with respect to 
both initial condition and $q$-TASEP dynamics.)
This looks very similar to (\ref{mom_step}) but there is an important difference. 
As can be understood by considering $t=0$ case of this formula, 
the moment is finite only for small values of $n$, satisfying 
$\a q^{-n}<\max_{m=1,\ldots , N}{a}_m$,
except for the step initial condition $\a=0$, and the higher moments diverge. 
Hence one can not use (\ref{qLapgen}) to calculate the $q$-Laplace transform. 
On the other hand, one has, by definition,  
\begin{equation}
\left\langle \frac{1}{(\zeta q^{x_N(t)+N};q)_\infty} \right\rangle 
= 
\sum_{l\in\mathbb{Z}} \frac{1}{(\zeta q^l;q)_{\infty}} \mathbb{P}[x_N(t)+N=l], 
\label{qLapdef}
\end{equation}
for $\zeta\neq q^n,n\in\mathbb{Z}$ and, since 
$\lim_{l\to-\infty}(\zeta q^l;q)_{\infty}=\infty, \lim_{l\to\infty}(\zeta q^l;q)_{\infty}=1$,
this $q$-Laplace transform is finite for the random initial condition (\ref{rand}) as well. 
One has to find a way to calculate the $q$-Laplace transform without using the $q$-moments. 

In \cite{IS2017p}, we have overcome this difficulty  and found a Fredholm determinant representation 
for the $q$-Laplace transform by using Ramanujan's summation formula and the Cauchy determinant
for theta functions (also known as the Frobenius determinant). 
In this note, we study the same problem and present a somewhat different approach without using them and 
give a different Fredholm determinant representation of the $q$-Laplace transform~\eqref{qLapgen} for the 
$q$-TASEP with the half stationary initial condition~\eqref{rand}. Our main result is the following. 
  
\begin{theorem}
For the $q$-TASEP with the random initial condition (\ref{rand}),(\ref{16}) we have 
\begin{align}
&\left\langle\frac{1}{(\zeta q^{x_N(t)+N};q)_{\infty}}\right\rangle
=
\det\left(
1
+
K_{\zeta}
\right)_{L^2(C_a)},
\label{qLapFr}
\end{align}
where $\zeta \neq q^n, n\in\mathbb{Z}$, $C_a$ is a contour around $a_i$'s 
and the kernel is given by 
\begin{align}
K_{\zeta}(w_1,w_2)
=
\frac{-1}{2\pi i}
\int_{i\R+\e} ds
\frac{(-\zeta)^s\pi}{\sin\pi s}
\frac
{e^{(q^{s}w_1-w_2)t}}
{q^{s}w_1-w_2}
\frac{(\a/w_1;q)_{\infty}}{(\a/(q^s w_1);q)_{\infty}}
\prod_{m=1}^N
\frac
{(q^{s}w_1/a_m;q)_{\infty}}
{(w_1/a_m;q)_{\infty}}.
\label{c121}
\end{align}
\end{theorem}
Once this type of formula is found, then there is a standard way to study the limiting distribution. 
In particular by setting $a_i=1,i\geq 2$ and taking $a_1\to \alpha$ limit and applying similar 
arguments as in \cite{IS2017p}, one can study the stationary $q$-TASEP and reproduce the 
results there that the limiting distribution of a particle is given by the Baik-Rains distribution. 
Compared to the kernel found in \cite{IS2017p}, the above kernel is much closer 
to the one in \cite{BC2014,BCS2014}. But we stress again that the method to find a kernel 
through the $q$-moments in \cite{BC2014,BCS2014}  does not work for the random initial 
condition (\ref{rand}). For a certain part, our approach has a similarity to the one 
in~\cite{BCR2013} for the log-Gamma and the O\rq{}Connell-Yor 
polymers (for the case corresponding to the step case).

For the case of ASEP, a somewhat different approach was employed to study the stationary case
in \cite{Aggarwal2016p}. There the author uses the fact that the ASEP can be obtained as a limit of 
the higher spin vertex model, because at the level of higher spin vertex model everything is discrete and 
all moments are finite. However, we emphasize that our approach, initiated in \cite{IS2017p}
for $q$-TASEP and generalized to the higher spin vertex model in \cite{IMuS2018p}, 
has an advantage that all models in the hierarchy can be treated directly in the same 
manner without relying on a limiting procedure. 
It would be an interesting question to clarify interrelationships between various 
approaches and representations. 

This note is organized as follows. In Sec.~\ref{sec:dist}, we recall a few facts 
from \cite{IS2017p}. We introduce a two-sided version of the $q$-Whittaker measure
and gives an multiple integral formula for the distribution of the position of a particle 
of the $q$-TASEP with the half-stationary initial condition ~\eqref{rand}. 
In Sec.~\ref{sec:qLap}, we explain a way to calculate the $q$-Laplace transform~\eqref{qLapgen}
and present a multiple integral formula for it. 
In Sec.~\ref{sec:det}, we obtain a Fredholm determinant representation and show that it is 
equivalen to~\eqref{qLapFr}

{\bf Notation}: In order to make some formulas look better, we replace $a_i$ by $q^{a_i}$
and $\a$ by $q^{\a}$ in the following sections.

\section{$q$-Whittaker measure and distribution of a particle position}
\label{sec:dist}
First we recall a few facts from \cite{IS2017p}. For a $N\in\mathbb{Z}_{>0}$, let $\Lambda_N$ be the set of 
signatures (or integer partitions) 
defined by
%
%
%
$\Lambda_N=\{\l=(\l_1,\cdots,\l_N)\in\Z^N|
\l_j\in\Z~\text{for} ~j=1,2,\cdots,N,
\l_1\ge\cdots\ge\l_N\}$.
Note that, unlike for the the case of usual partitions,  each $\l_j$ can be a negative integer. 
Let $a_i\geq 0,\a_i\geq 0,1\leq i\leq N$ and 
$q^c=(q^{c_1},\ldots,q^{c_N})$ for $c=a,\a$. 
We introduce a measure on $\Lambda_N$, 
\begin{align}
W_t(\l)
=\frac{P_\l(q^a)Q_\l(q^\a,t)}{\Pi (q^a;q^\a,t)},
\label{ptn}
\end{align}
where $P_{\l},Q_{\l}$ are the $q$-Whittaker functions defined by 
\begin{align}
P_\l(q^a) 
&=  
\sum_{\substack{\l_i^{(j)}\in\mathbb{Z}, 1\leq i\leq j\leq N-1\\
            \l_{i+1}^{(j+1)} \leq \l_i^{(j)} \leq \l_i^{(j+1)}\\
            \l_i^{(N)}=\l_i}
            }
	\prod_{j=1}^N
	\prod_{i=1}^N q^{a_j {\l^{(j)}_i}}\cdot
	\prod_{i=1}^{N-1}\frac{ q^{-a_j \l^{(j-1)}_i}(q;q)_{\l_i^{(j)}-\l_{i+1}^{(j)}}}
        {(q;q)_{\l_i^{(j)}-\l_i^{(j-1)}}(q;q)_{\l_i^{(j-1)}-\l_{i+1}^{(j)}}}, \label{P}\\
Q_\l(q^a) &= \prod_{i=1}^{N-1}(q^{\l_i-\l_{i+1}+1};q)_{\infty}
\int_{\T^N}\prod_{i=1}^N\frac{dz_i}{z_i}\cdot P_{\l}(1/z)
 \Pi\left(z;q^\a,t\right)m_N^q\left(z\right)
\label{Q}
\end{align}
with 
\begin{equation}
m_N^q(z)=\frac{1}{(2\pi i)^NN!}\prod_{1\le i<j\le N}(z_i/z_j;q)_{\infty}(z_j/z_i;q)_{\infty}
\label{qsk}
\end{equation}
and 
\begin{equation}
\Pi\left(q^a;q^\a,t\right)
=
\prod_{i,j=1}^N\frac{1}{(q^{\a_i-a_j};q)_{\infty}}\cdot\prod_{j=1}^Ne^{q^{a_j}t} .
\label{Pi}
\end{equation}
The measure (\ref{ptn}) is called the two-sided $q$-Whittaker measure. 

Let $P_t(\l_N)$ denote the marginal distribution of $\l_N$ under the above two-sided $q$-Whittaker measure,
$P_t(\l_N)= \sum_{\l_i,1\leq i\leq N-1} W_t(\l)$. 
By rewriting the Cauchy identity for the $q$-Whittaker function, 
we found the following multiple integral formula \cite{IS2017p}. 
\begin{proposition}\label{p8}
The marginal distribution of $\l_N$ under the two-sided $q$-Whittaker measure (\ref{ptn}) is given by  
\begin{align}
P_t(\l_N)=(q;q)_{\infty}^{N-1}
\int_{\T^N}\prod_{j=1}^N \frac{dz_j}{z_j}
\cdot
\left(
\frac{q^{A}}{z_1\cdots z_N}
\right)^{\lambda_N}
m^q_N(z)
\frac{\Pi(z;q^\a,t)}{\Pi(q^a;q^\a,t)}
\cdot
\frac
{\left(q^{A}/z_1\cdots z_N;q\right)_{\infty}}
{\prod_{i,j=1}^N(q^{a_i}/z_j;q)_{\infty}}
\label{p82}
\end{align}
with $A=\sum_{j=1}^n a_j$. 
\end{proposition}
\noindent
Note that on the right hand side the dependence on $\l_N$ appears only as the power 
of a factor in the integrand. This is useful for further calculations. 

As shown in \cite{IS2017p}, we have, with $\a_1=\a, \a_i\to\infty, i\geq 2$,
\begin{equation}
 \mathbb{P}[x_N(t)+N=\l_N] = P_t(\l_N),
 \label{xNlambda}
\end{equation} 
that is, the probability $\mathbb{P}[x_N(t)+N=\l_N],\l_N\in\mathbb{Z}$ that 
the position of the $N$-th particle in $q$-TASEP with the half-stationary initial condition (\ref{rand})
is $\l_N$ is the same as the marginal distribution of $\lambda_N$ for the two 
sided $q$-Whittaker measure with $\a_1=\a, \a_i\to\infty, i\geq 2$. In the sequel we focus on the 
study of $P_t(\l_N)$ for general $a_i$'s and $\a_i$'s.

\section{$q$-Laplace transform}
\label{sec:qLap}
In this section we consider the $q$-Laplace transform of
$P_t(\l_N)$. Our strategy is to evaluate it directly without using
the $q$-moments~\eqref{mom_rand}. First we obtain the 
following integral representation.
\begin{proposition}\label{p6}
For $\e>0$ and $\zeta\in\C\setminus\R_+$, we have
\begin{align}
\left\langle\frac{1}{(\zeta q^{\l_N};q)_{\infty}}\right\rangle
&=
-(q;q)_{\infty}^{N-2}
\int_{i\R -\e}ds
\frac{\pi}{\sin \pi(s-A)}
(-\zeta)^{s-A}(q^{s-A+1};q)_{\infty}\left(q^{A-s};q\right)_{\infty}
\notag
\\
&~~~~\times\int_{\T^{N-1}}\prod_{j=2}^N\frac{dw_j}{w_j}
\cdot
m^q_N(w)\frac{\Pi(w;q^\a,t)}{\Pi(q^a;q^\a,t)}
\prod_{i,j=1}^N\frac{1}{\left(\frac{q^{a_i}}{w_j};q\right)_{\infty}},
\label{p60}
\end{align}
where $m_N^q(w)$, $\Pi(w;q^{\a},t)$ and $A$ are defined by~\eqref{qsk},~\eqref{Pi}
and below~\eqref{p82} respectively and $w_1$ is defined by
\begin{align}
w_1
=
\begin{cases}
q^s,& \text{for~} N=1,\\
q^{s}/w_2\cdots w_N,& \text{for~} N\ge 2.
\end{cases}
\label{p900}
\end{align}
\end{proposition}
\smallskip
\noindent
{\bf Proof.}
In this proof we assume $|\zeta|<1$. Once (\ref{p60}) is proved for $|\zeta|<1$, the extension to 
$\zeta\in\C\setminus\R_+$ is easy by analytic continuation. 
We start from writing LHS of~\eqref{p60} as
\begin{align}
\sum_{\l_N=-\infty}^{\infty}
\frac{1}{(\zeta q^{\l_N};q)_{\infty}}P_t\left(\l_N\right),
\label{p91}
\end{align}
where $P_t\left(\l_N\right)$ is given by~\eqref{p82}. 
(This is the same as (\ref{qLapdef}) by (\ref{xNlambda}).) 

Next we separate the sum into two parts, $0\le \l_N$(the positive part) and 
$\l_N\le -1$(the negative part). The calculation of the positive part is straightforward.
Since we assumed $|\zeta|<1$, by an application of the $q$-binomial theorem, it is written as 
\begin{equation}
%
\sum_{n=0}^{\infty}\int_{\mathbb{T}^N}
\frac{\zeta^n}{(q;q)_n}
\prod_{j=1}^N dz_j
\cdot
\frac{m^q_N(z_1,\cdots,z_N)}{z_1\cdots z_n-q^{A+n}}
\frac{\Pi(z;q^\a,t)}{\Pi(q^a;q^\a,t)}
\cdot
\frac{\left(\frac{q^A}{z_1\cdots z_N};q\right)_{\infty}(q;q)_{\infty}^{N-1}}{\prod_{i,j=1}^N(\frac{q^{a_i}}{z_j};q)_{\infty}}.
\label{p65}
\end{equation}
For the negative part, we will show that it is written as 
\begin{align}
&-
\sum_{n=0}^{\infty}\int_{\mathbb{T}^N}
\prod_{j=1}^N dz_j
\cdot
\frac{\zeta^n}{(q;q)_n}
\frac{m^q_N(z_1,\cdots,z_N)}{z_1\cdots z_n-q^{A+n}}
\frac{\Pi(z;q^\a,t)}{\Pi(q^a;q^\a,t)}
\cdot
\frac{\left(\frac{q^A}{z_1\cdots z_N};q\right)_{\infty}(q;q)_{\infty}^{N-1}}{\prod_{i,j=1}^N(\frac{q^{a_i}}{z_j};q)_{\infty}}
\notag\\
&-(q;q)_{\infty}^{N-2}
\int_{i\R -\e}ds
\frac{\pi}{\sin \pi(s-A)}
(-\xi)^{s-A}(q^{s-A+1};q)_{\infty}\left(q^{A-s};q\right)_{\infty}
\notag
\\
&\hspace{3.5cm}\times\int_{\T^{N-1}}\prod_{j=2}^N\frac{dw_j}{w_j}
\cdot
m^q_N(w)\frac{\Pi(w;q^\a,t)}{\Pi(q^a;q^\a,t)}
\prod_{i,j=1}^N\frac{1}{\left(\frac{q^{a_i}}{w_j};q\right)_{\infty}},
\label{p62}
\end{align}
where $w_1$ in the second term is defined by~\eqref{p900}. Combining
\eqref{p91}--\eqref{p62}, we obtain our desired result~\eqref{p60}.

Note that the $q$-binomial theorem is not applicable to the negative part since 
$|\zeta q^{\l_N}|$ becomes bigger than one as $\l_N\to -\infty$.  
Let us introduce $L\in\Z_{> 0}$ and $x\in\C$,  
and consider the quantity,
\begin{align}
\sum_{\ell=-L}^{-1}
\frac{1}{(\zeta x^{-\ell};q)_{\infty}}
P_t\left(\ell\right).
\label{p610}
\end{align}
As a function of $x$, this is analytic for $|x|<1$. 
Below we will extend  this region of analyticity in $x$, 
$|x|<1$, to one which includes $x=1/q$ and then take the limit $L\rightarrow\i$.
Note that the case $x=q^{-1}$ with this limit corresponds to LHS of~\eqref{p62}.
Note that, when $|x|<1$,  the $q$-binomial theorem is applicable
for arbitrary fixed $L$ since $|\zeta x^L|<1$ is satisfied.  
Hence
\eqref{p610} can be written easily as
\begin{align}
\sum_{n=0}^\infty \frac{\zeta^n}{(q;q)_n}\int_{\T^N}\prod_{j=1}^N\frac{dz_j}{z_j}
\cdot
\sum_{\ell=1}^L\left(\frac{x^n z_1\cdots z_N}{q^A}\right)^{\ell}
\cdot
 m^q_N(z)\frac{\Pi(z;q^\a,t)}{\Pi(q^a;q^\a,t)}
\cdot
\frac{\left(\frac{q^A}{z_1\cdots z_N};q\right)_{\infty}(q;q)_{\infty}^{N-1}}{\prod_{i,j=1}^N(\frac{q^{a_i}}{z_j};q)_{\infty}}.
\label{p611}
\end{align}
Then we rewrite the sum over $n$ as the contour integral by the residue theorem as 
\begin{align}
\sum_{n=0}^\infty \frac{\zeta^n x^{\ell n}}{(q;q)_n} 
=
\frac{-1}{2 \pi i}\int_{i\R-\e}du
\frac{\pi}{\sin\pi u} 
\frac{(-\zeta x^{\ell})^u(q^{1+u};q)_{\i}}{(q;q)_{\i}},
\end{align}
where we set the branch cut of the function $z^u$ as $\R_-$
(see Lemma 3.20 and Step 2 in the proof of Theorem 3.18 in~\cite{BC2014} for a similar identity). 
We find
\begin{align}
&\sum_{\ell=-L}^{-1}
\frac{1}{(\zeta x^{-\ell};q)_{\infty}}
P_t\left(\ell\right)
=
\int_{i\R -\e}\frac{du}{2\pi i}
\frac{\pi}{\sin \pi u}
\frac{(-\zeta)^u(q^{u+1};q)_{\infty}}{(q;q)_{\infty}}
\notag\\
&\times
\int_{\T^N}\prod_{j=1}^N \frac{dz_j}{z_j}
\cdot
\sum_{\ell=1}^L\left(\frac{x^u z_1\cdots z_N}{q^A}\right)^{\ell}
\cdot
\frac{\left(\frac{q^A}{z_1\cdots z_N};q\right)_{\infty}(q;q)_{\infty}^{N-1}}{\prod_{i,j=1}^N(\frac{q^{a_i}}{z_j};q)_{\infty}}
\label{p612}
\end{align}
with $\e>0$.
Now we consider the analytic continuation for $x$ in both hand sides 
of~\eqref{p612}. Let $\mu, \theta\in(-\pi,\pi)$ be 
the argument of $-\zeta$ and 
$x$ respectively, i.e., $-\zeta=|\zeta|e^{i\mu},~x=|x|e^{i\th}$.
In lhs, one can see that it is analytic for $x\in\C\setminus\Omega$
where
\begin{align}
\Omega=
\left\{
\frac{e^{i(\mu+\pi)/\ell}}{|\zeta|q^n}
;\ell\in\mathbb{Z}_{>0},L, n\in\mathbb{N} 
\right\} .
\end{align}
Thus we see that by analytic continuation one can set $x=1/q$ since,
for $\mu\in (-\pi,\pi)$, one finds $1/q\notin\Omega$.

In rhs of~\eqref{p612}, let us focus on the integral
\begin{align}
\int_{i\R-\e}\frac{du}{2\pi i}
\frac{\pi}{\sin\pi u} (-\zeta x^{\ell})^u(q^{u+1};q)_{\infty} 
\label{p615}
\end{align}
for $\ell=1,2,\cdots,L$. Considering the fact that $|q^{ix}|=1$ for $x\in\R$
and
\begin{align}
\lim_{x\rightarrow\pm \infty}\frac{e^{-\pi |x|}}{2 i \sin(\pi i x)}=1,
\end{align}
we find that~\eqref{p615} is analytic as a function of $x$ if 
\begin{align}
\lim_{y\rightarrow\pm\infty}
\frac{(-\zeta x^\ell)^{iy}}{e^{\pi|y|}}=0
\Leftrightarrow
\frac{-\pi-\mu}{\ell}<\theta<\frac{\pi-\mu}{\ell}
\label{p616}
\end{align}
for $\ell=1,2,\cdots,L$. 
Note that the most strict case $\ell=L$,
\begin{align}
\frac{-\pi-\mu}{L}<\theta<\frac{\pi-\mu}{L}
\label{p617}
\end{align}
includes the vicinity of $\theta=0$ for any fixed $\mu\in (-\pi,\pi)$.
Thus under the condition~\eqref{p617},
we can extend the region of analyticity from $|x|<1$ to a larger region including $\mathbb{R}_+$
and in particular we can set $x=1/q$.

Thus setting $x=1/q$ and taking the sum for $\ell$, 
we get
%
%
\begin{align}
&\sum_{\ell=-L}^{-1}
\frac{1}{(\zeta q^{\ell};q)_{\infty}}
P_t\left(\ell\right)
=(q;q)_{\infty}^{N-2}
\int_{i\R -\e}\frac{du}{2\pi i}
\frac{\pi}{\sin \pi u}
(-\zeta)^u(q^{u+1};q)_{\infty}
\notag\\
&\times
\int_{\T^N}\prod_{j=1}^N dz_j
\cdot
\frac{1-(z_1\cdots z_N/q^{A+u})^L}{z_1\cdots z_N-q^{A+u}}
m^q_N(z)\frac{\Pi(z;q^\a,t)}{\Pi(q^a;q^\a,t)}
\cdot
\frac{\left(\frac{q^A}{z_1\cdots z_N};q\right)_{\infty}}{\prod_{i,j=1}^N(\frac{q^{a_i}}{z_j};q)_{\infty}}.
\label{p612a}
\end{align}
Note that at this stage we can take $\e$ to be an arbitrary positive
real value. Here we set it to be $\e_A$ such that $\e_A>A$, which leads to 
$|z_1\cdots z_N/q^{A+u}|<1$. Then we can take the $L\rightarrow\infty$
limit and have
\begin{align}
\sum_{\ell=-\infty}^{-1}
\frac{1}{(\zeta q^{\ell};q)_{\infty}}
P_t\left(\ell\right)
&=(q;q)_{\infty}^{N-2}
\int_{i\R -\e_A}\frac{du}{2\pi i}
\frac{\pi}{\sin \pi u}
(-\zeta)^u(q^{u+1};q)_{\infty}
\notag\\
&\times
\int_{\T^N}\prod_{j=1}^N dz_j
\cdot
\frac{1}{z_1\cdots z_N-q^{A+u}}
\cdot
 m^q_N(z)\frac{\Pi(z;q^\a,t)}{\Pi(q^a;q^\a,t)}
\cdot
\frac{\left(\frac{q^A}{z_1\cdots z_N};q\right)_{\infty}}{\prod_{i,j=1}^N(\frac{q^{a_i}}{z_j};q)_{\infty}}.
\end{align}
For later use, we change the variables $w_1=z_1\cdots z_N$ and $w_j=z_j$ for
$j\ge 2$. We have
\begin{align}
&\sum_{\ell=-\infty}^{-1}
\frac{1}{(\zeta q^{\ell};q)_{\infty}}
P_t\left(\ell\right)
\notag\\
&=(q;q)_{\infty}^{N-2}
\int_{i\R -\e_A}\frac{du}{2\pi i}
\frac{\pi}{\sin \pi u}
(-\zeta)^u(q^{u+1};q)_{\infty}
\int_{\T^{N-1}}\prod_{j=2}^N\frac{dw_j}{w_j}
\cdot
\int_\T dw_1\frac{ C(w;a,\a)}{w_1-q^{A+u}},
\end{align}
where 
\begin{align}
C(w;a,\a)=
\left.
m^q_N(z)\frac{\Pi(z;q^\a,t)}{\Pi(q^a;q^\a,t)}
\cdot
\frac{\left(\frac{q^A}{z_1\cdots z_N};q\right)_{\infty}}{\prod_{i,j=1}^N(\frac{q^{a_i}}{z_j};q)_{\infty}}
\right|_{z_1=w_1/w_2\cdots w_N,~z_j=w_j~\text{for}~j\ge 2}.
\end{align}

We change the contour $\T$ of $w_1$ to $\T_A$ such that the contour
encloses the pole $w_1=q^{A+u}$. We have 
\begin{align}
&~\sum_{\ell=-\infty}^{-1}
\frac{1}{(\zeta q^{\ell};q)_{\infty}}
P_t\left(\ell\right)
\notag\\
&=(q;q)_{\infty}^{N-2}
\int_{i\R -\e_A}\frac{du}{2\pi i}
\frac{\pi}{\sin \pi u}
(-\zeta)^u(q^{u+1};q)_{\infty}
\int_{\T^{N-1}}\prod_{j=2}^N\frac{dw_j}{w_j}
\cdot
\int_{\T_A} dw_1\frac{C(w;a,\a)}{w_1-q^{A+u}}
\notag\\
&-(q;q)_{\infty}^{N-2}
\int_{i\R -\e_A}\frac{du}{2\pi i}
\frac{\pi}{\sin \pi u}
(-\zeta)^u(q^{u+1};q)_{\infty}
\int_{\T^{N-1}}\prod_{j=2}^N\frac{dw_j}{w_j}
\cdot
\underset{w_1=q^{A+u}}{\text{Res}}
\frac{C(w;a,\a)}{w_1-q^{A+u}}.
\label{p618}
\end{align}
Replacing the integration of $u$ by the summation of
the residues at $u=0,1,2,\cdots$ and changing the variables $w_j$ to $z_j$, 
we find that the first term of~\eqref{p618} becomes
\begin{align}
-\sum_{n=0}^{\infty}\int_{\mathbb{T}^N}
\prod_{j=1}^N dz_j
\cdot
\frac{\zeta^n}{(q;q)_n}
\frac{m^q_N(z_1,\cdots,z_N)}{z_1\cdots z_n-q^{A+n}}
\frac{\Pi(z;q^\a,t)}{\Pi(q^a;q^\a,t)}
\cdot
\frac{\left(\frac{q^A}{z_1\cdots z_N};q\right)_{\infty}}{\prod_{i,j=1}^N(\frac{q^{a_i}}{z_j};q)_{\infty}},
\end{align}
which is exactly equal to the first term in~\eqref{p62} and cancels~\eqref{p65}. 
We easily find that the second term
is written as
\begin{align}
-(q;q)_{\infty}^{N-2}\int_{i\R -\e_A}\frac{du}{2\pi i}
\frac{\pi}{\sin \pi u}
(-\xi)^u(q^{u+1};q)_{\infty}
\int_{\T^{N-1}}\prod_{j=2}^N\frac{dw_j}{w_j}
\cdot
m^q_N(w)\frac{\Pi(w;q^\a,t)}{\Pi(q^a;q^\a,t)}
\frac{\left(q^{-u};q\right)_{\infty}}{\prod_{i,j=1}^N(\frac{q^{a_i}}{w_j};q)_{\infty}}
\end{align}
with $w_1=q^{A+u}/w_2\cdots w_N$. Shifting $u$ as $s=u+A$, we arrive at
the second term in~\eqref{p62}. 
\qed

Now we have another integral representation, which is more useful
for our purpose. 
\begin{proposition}
\label{p10}
For $\e>0$ and $\zeta\in\C\setminus\R_+$, we have
\begin{align}
\left\langle\frac{1}{(\zeta q^{\l_N};q)_{\infty}}\right\rangle
&=
\frac{(-\pi)^N}{(2\pi i)^NN!}
\int_{(i\R-\e)^N}
\prod_{j=1}^N \frac{ds_j}{q^{(N-1)a_j}(q;q)_{\infty}}
	\frac{(-\zeta)^{s_j}}{(-\zeta)^{a_j}}
\cdot
\frac
{\Pi(q^s;q^\a,t)}
{\Pi(q^a;q^\a,t)}
\cdot
\prod_{i,j=1}^N\frac{(q^{s_i-a_j+1};q)_{\infty}}{\sin\pi(s_i-a_j)}\notag\\
&
\hspace{1.4cm}
\times
\prod_{1\le i<j\le N}\frac{\sin\pi(s_j-s_i)\sin\pi(a_j-a_i)(q^{s_j}-q^{s_i})(q^{a_j}-q^{a_i})}
{(q^{a_i-a_j};q)_{\infty}(q^{a_j-a_i};q)_{\infty}}.
\label{p100}
\end{align}
\end{proposition}
\smallskip
\noindent
{\bf Proof.} 
As in the proof Proposition 3, we assume $|\zeta|<1$ in this proof. 
When $|\zeta|<1$, we find that both~\eqref{p60} and~\eqref{p100}
can be evaluated as sums of residues on the right-half plane
since both integrands vanishes as $\Re s\rightarrow\i$. 

We see that the integrand in~\eqref{p60} has poles at the following points: 
\begin{align}
&s=\a_k+\frac{\log\prod_{j=2}^Nw_j}{\log q}+n_1+\frac{2\pi i\ell}{\log q},~
a_k+\frac{\log\prod_{j=2}^Nw_j}{\log q}+n_1+\frac{2\pi i\ell}{\log q},~
\notag\\
&w_j=q^{\a_k+n_j},~q^{a_k+n_j},~j=2,3,\cdots,N
\label{p101}
\end{align}
for $k=1,2,\cdots,N$, $n_1,\cdots,n_N\in\N$ and $\ell\in\Z$.
By using $w_1$~\eqref{p900} in place of $s$, \eqref{p101} can be written 
in a more compact form,
\begin{align}
w_l=q^{\a_k+n_l},~q^{a_k+n_l},~l=1,\cdots,N, 
\label{p1002}
\end{align}
for $k=1,2,\cdots,N$ and $n_l\in\N$.
However we find that when at least one pair of $w_j$\rq{}s
shares common $a_k$ or $\a_k$, the choices have no contribution:
e.g. in both cases 
\begin{align}
(w_j,w_k)=(q^{a_m+n_j},~q^{a_m+n_k}),~(q^{\a_m+n_j},~q^{\a_m+n_k})
\end{align}
for some $j,k$ and $m$, we find that the factor $\left(w_j/w_k;q\right)_\infty\left(w_j/w_k;q\right)_\infty$ in $m_N^q(w)$
in the integrand of~\eqref{p60} vanishes,
\begin{align}
\left(w_j/w_k;q\right)_\infty\left(w_j/w_k;q\right)_\infty
=\left(q^{n_j-n_k};q\right)_\infty\left(q^{n_k-n_j};q\right)_\infty
=0.
\end{align}
Thus we need to consider only the poles where all $w_j~j=1,2,\cdots,N$ have
distinct $a_k$\rq{}s or $\a_k$\rq{}s.
Furthermore the pole contribution does not change under the exchange of
the poles since the integrand in~\eqref{p60} is symmetric under the exchange 
of $w_i$\rq{}s. Thus we find that~\eqref{p60} can be represented as follows:
\begin{align}
\left\langle\frac{1}{(\zeta q^{\l_N};q)_{\infty}}\right\rangle
=
N!\sum_{N_1=0}^{\infty}
&\sum_{1\le j_1<\cdots<j_{N_1}\le N}
\sum_{1\le j_{N_1+1}<\cdots<j_{N}\le N}
\notag
\\
&\sum_{n_1,\cdots,n_N=0}^{\infty}
\sum_{\ell\in\Z}
f(a_{j_1},\cdots,a_{j_{N_1}},\a_{j_{N_1+1}},\cdots ,\a_{j_N},{\bf n},\ell,N_1),
\label{p1004}
\end{align}
where ${\bf n}=(n_1,\cdots,n_N)$, and
$f(a_{j_1},\cdots,a_{j_{N_1}},\a_{j_{N_1+1}},\cdots ,\a_{j_N},{\bf n},\ell,N_1)
$ denotes the residue of the poles
\begin{align}
w_k=
\begin{cases}
q^{a_{j_k}+n_k},&~\text{for}~ k=1,\cdots,N_1,
\\
q^{\a_{j_k}+n_k},&~\text{for}~ k=N_1+1,\cdots,N.
\end{cases}
\end{align}
Note that in this case the corresponding pole for $s$ is at
\begin{align}
s=\sum_{k=1}^{N_1}a_{j_k}+\sum_{k=N_1+1}^N\a_{j_k}+\sum_{j=1}^Nn_j+\frac{2\pi i\ell}{\log q}. 
\end{align}
%
%

Similarly~\eqref{p100} has poles at
\begin{align}
s_j=a_k+n_{j},~\a_k+n_j+\frac{2\pi i\ell_k}{\log q}, 
\end{align}
where $j,k=1,2,\cdots N$,  $n_j\in \Z_{\ge 0}$ and $\ell_j\in\Z$. 
We easily find that when at least one pair of  $s_j, j=1,2\cdots,N$
shares the common $a_k$ or $\a_k$, they have no contribution: 
E.g. when $(s_i,s_j)=(a_k+n_i,~a_k+n_j)$, the factor $\sin \pi (s_j-s_i)$
in the numerator in~\eqref{p100} becomes
\begin{align}
\sin \pi (s_j-s_i)=\sin\pi (n_j-n_i)=0 .
\end{align}
Similarly we find that 
$(s_{j_1},s_{j_2})=(\a_k+n_{j_1}+ 2\pi i\ell_{j_1}/\log q,~\a_k+n_{j_2}+2\pi i\log \ell_{j_2}/\log q)$ cancels
the contribution $(s_{j_1},s_{j_2})=(\a_k+n_{j_1}+ 2\pi i\ell_{j_2}/\log q,~\a_k+n_{j_2}+
2\pi i\ell_{j_1}/\log q)$ 
since in the former case, the same factor $\sin \pi (s_{j_2}-s_{j_1})$ produces 
\begin{align}
\sin\pi (s_{j_2}-s_{j_1})=(-1)^{n_{j_2}-n_{j_1}}\sin\left(\frac{2\pi i (\ell_{j_2}-\ell_{j_1})}{\log q}\right)
\end{align}
and in the latter one, it gives the same quantity with opposite sign.
Furthermore the residue of the poles with distnct $a_k$\rq{}s or $\a_k$\rq{}s
does not change under the exchange of the poles since the integrand in~\eqref{p100}
is symmetric under the exchange of $s_j$\rq{}s. Thus we arrive at the following
expression for~\eqref{p60}
\begin{align}
\left\langle\frac{1}{(\zeta q^{\l_N};q)_{\infty}}\right\rangle
&=
N!\sum_{N_1=0}^{\infty}
\sum_{1\le j_1<\cdots<j_{N_1}\le N}
\sum_{1\le j_{N_1+1}<\cdots<j_{N}\le N}
\notag
\\
&~\times\sum_{n_1,\cdots,n_N=0}^{\infty}
\sum_{\ell_{N_1+1},\cdots,\ell_N\in\Z}
g(a_{j_1},\cdots,a_{j_{N_1}},\a_{j_{N_1+1}},\cdots,\a_{j_N},{\bf n},\bl,N_1).
\label{p1003}
\end{align}
where ${\bf n}=(n_1,\cdots,n_N)$, $\bl=(\ell_{N_1+1},\cdots,\ell_N)$,
and
$
g(a_{j_1},\cdots,a_{j_{N_1}},\a_{j_{N_1+1}},\cdots,\a_{j_N},{\bf n},\bl,N_1)
$
represents the residue at the poles,
\begin{align}
s_k=
\begin{cases}
a_{j_k}+n_k,& k=1,2,\cdots,N_1,
\\
\a_{j_k}+n_{k}+\frac{2\pi i\ell_k}{\log q}, & k=N_1+1,\cdots,N.
\end{cases}
\end{align}

In Appendix~\ref{a}, we will prove
\begin{align}
f(a_{j_1},\cdots,a_{j_{N_1}},\a_{j_{N_1+1}},\cdots \a_{j_N},{\bf n},\ell,N_1)
=\hspace{-5mm}
\sum_{\substack{\ell_{N_1+1},\cdots,\ell_N\in\Z\\
\ell_{N_1+1}+\cdots+\ell_N=\ell
}}
\hspace{-2mm}
g(a_{j_1},\cdots,a_{j_{N_1}},\a_{j_{N_1+1}},\cdots,\a_{j_N},{\bf n},\bl,N_1),
\label{p105}
\end{align}
from which the equivalence of the two expressions~\eqref{p1004} 
and~\eqref{p1003} immediately follows.

\qed


\section{Fredholm determinant formulas}
\label{sec:det}

In this section, we obtain a Fredholm determinant representation for the $q$-Laplace transform 
$\left\langle\frac{1}{(\zeta q^{\l_N};q)_{\infty}}\right\rangle$.
For this purpose we use Proposition~\ref{p6} and the following rational and trigonometric 
versions of the Cauchy identities, 
\begin{align}
&
\frac{
\prod_{1\le i<j\le N}
(q^{a_i}-q^{a_j})
(q^{s_j}-q^{s_i})
}
{
\prod_{i,j=1}^N
(q^{s_i}-q^{a_j})
}
=
\det
\left(
\frac{1}
{q^{s_i}-q^{a_j}}
\right)_{i,j=1}^N,
\label{41}
\\
&
\frac{
\prod_{1\le i<j\le N}
\sin\pi (a_i-a_j)
\sin\pi(s_j-a_i)
}
{
\prod_{i,j=1}^N
\sin\pi(s_i-a_j)
}
=
\det
\left(
\frac{1}
{\sin\pi(s_i-a_j)}
\right)_{i,j=1}^N.
\label{42}
\end{align}
For more general description about the Cauchy determinants, see for instance~\cite{KN2003}. 

\begin{theorem}
For the two-sided $q$-Whittaker measure (\ref{ptn}), we have, for $\zeta\neq q^n,n\in\mathbb{Z}$, 
\begin{align}
\left\langle\frac{1}{(\zeta q^{\l_N};q)_{\infty}}\right\rangle
=
\det\left(1-fK\right)_{L^2(\R)}.
\label{t110}
\end{align}
Here the kernel is defined by
\begin{align}
&f(x)=\frac{-\zeta}{-\zeta+e^x},
\label{t1101}
\\
&K(x_1,x_2)
=\sum_{k=0}^{N-1}\phi_k(x_1;a,\a,t)\psi_{k}(x_2;a,\a,t),
\label{t1102}
\end{align}
and the functions $\phi_k(x_1;a,\a,t)$ and $\psi_k(x_2;a,\a,t)$ are
given as
\begin{align}
&\phi_k(x_1;a,\a,t)
=
\sqrt{q^{a_{k+1}}-q^{\a_{k+1}}}
\int_D\frac{\log q~dv}{2\pi i}
\frac{e^{v x_1+q^v t}}
{q^{(N-1)v}(q^{v}-q^{a_{k+1}})}
\prod_{\ell=1}^k
\frac{q^v-q^{\a_\ell}}
{q^{v}-q^{a_\ell}}
\cdot
\prod_{m=1}^N
\frac{(q^{s-\a_m+1};q)_{\infty}}
{(q^{a_m-s+1};q)_{\infty}},
\label{t1103}
\\
&\psi_k(x_2;a,\a,t)
=
\sqrt{q^{a_{k+1}}-q^{\a_{k+1}}}
\int_{i\R}\frac{ds}{2\pi i}
\frac{e^{s x_2+q^s t-\delta |s|}q^{Ns}}
{q^{s}-q^{\a_{k+1}}}
\prod_{\ell=1}^k
\frac{q^s-q^{a_\ell}}
{q^s-q^{\a_\ell}}
\cdot
\prod_{m=1}^N
\frac{(q^{s-a_m+1};q)_{\infty}}
{(q^{\a_m-s+1};q)_{\infty}},
\label{t1104}
\end{align}
where in~\eqref{t1103} the convergence factor $\delta=0+$ is introduced and
$D$ denotes the contour enclosing $a_j,~j=1,\cdots,N$ positively.
\end{theorem}

\smallskip

\noindent
{\bf Proof.}
We will show that $N$-fold integral representation~\eqref{p100} can be rewritten as~\eqref{t110}.  
The following calculations have certain similarities to the ones in ~\cite{BCR2013}
for the O'Connell-Yor and the $\log$-Gamma polymers model.  

Note that the factor in~\eqref{p100} 
\begin{align}
\prod_{i,j=1}^N\frac{(q^{s_i-a_j+1};q)_{\infty}}{\sin\pi(s_i-a_j)}
\prod_{1\le i<j\le N}\frac{\sin\pi(s_j-s_i)\sin\pi(a_j-a_i)(q^{s_j}-q^{s_i})(q^{a_j}-q^{a_i})}
{(q^{a_i-a_j};q)_{\infty}(q^{a_j-a_i};q)_{\infty}}
\end{align}
can be written as
\begin{align}
&(-1)^N
\prod^N_{
\substack{
ij=1 \\
i\neq j
}
}
\frac{1}{(q^{a_i-a_j};q)_{\infty}}
\prod_{i,j=1}^N (q^{s_i-a_j};q)_{\infty}
\cdot
\prod_{j=1}^Nq^{N_{a_j}}
\notag
\\
&\hspace{3cm}\times
\frac{
\prod_{1\le i<j\le N}
\sin\pi (a_i-a_j)
\sin\pi(s_j-a_i)
}
{
\prod_{i,j=1}^N
\sin\pi(s_i-a_j)
}
\cdot
\frac{
\prod_{1\le i<j\le N}
(q^{a_i}-q^{a_j})
(q^{s_j}-q^{s_i})
}
{
\prod_{i,j=1}^N
(q^{s_i}-q^{a_j})
}
\notag\\
&=
(-1)^N
\prod^N_{
\substack{
ij=1 \\
i\neq j
}
}
\frac{1}{(q^{a_i-a_j};q)_{\infty}}
\prod_{i,j=1}^N (q^{s_i-a_j};q)_{\infty}
\cdot
\prod_{j=1}^Nq^{N{a_j}}
\notag
\\
&\hspace{5cm}\times
\det
\left(
\frac{1}
{q^{s_i}-q^{a_j}}
\right)_{i,j=1}^N
\cdot
\det
\left(
\frac{1}
{\sin\pi(s_i-a_j)}
\right)_{i,j=1}^N,
\label{t111}
\end{align}
where in the last expression we used~\eqref{41} and~\eqref{42}.
Substituting this into~\eqref{p100}, we get
\begin{align} 
\left\langle\frac{1}{(\zeta q^{\l^{(N)}_N};q)_{\infty}}\right\rangle
&=
\frac{1}{(2i)^NN!}
\int_{(i\R-\e)^N}
\prod_{j=1}^N \frac{ds_j q^{a_j}}{(q;q)_{\infty}}
	\frac{(-\zeta)^{s_j}\Pi(q^{s_j};q^{\a},t)}{(-\zeta)^{a_j}\Pi(q^{a_j};q^\a,t)}
\cdot
\frac
{\prod_{i,j=1}^N(q^{s_i-a_j};q)_{\infty}}
{\prod_{\substack{i,j=1\\ i\neq j}}^N (q^{a_j-a_i};q)_{\infty}}
\notag
\\
&\hspace{1cm}
\times
\det
\left(
\frac{1}
{q^{s_i}-q^{a_j}}
\right)_{i,j=1}^N
\cdot
\det
\left(
\frac{1}
{\sin\pi(s_i-a_j)}
\right)_{i,j=1}^N.
\label{t112}
\end{align}

Once we get this type of determinantal formula, it is straightforward to reach 
the Fredholm determinant.
Noting that Andr{\'e}ief identity can be applicable to~\eqref{t112},
we get a single determinant expression with rank $N$,
\begin{align}
&\left\langle\frac{1}{(\zeta q^{\l^{(N)}_N};q)_{\infty}}\right\rangle
\notag\\
&=
\det\left(
\frac{1}{2\pi i}
\int_{i\R-\e}
\frac{ds q^{a_j}}{(q;q)_{\infty}}
	\frac{(-\zeta)^{s}\Pi(q^s;q^\a,t)}{(-\zeta)^{a_j}\Pi(q^{a_j};q^\a,t)}
\cdot
\frac{\prod_{m=1}^N(q^{s-a_m};q)_{\infty}}{\prod_{\substack{m=1\\ m\neq j}}^N
(q^{a_j-a_m};q)_{\infty}}
\frac{q^{a_i}}
{q^s-q^{a_i}}
\frac{\pi}{\sin\pi(s-a_j)}
\right)_{i,j=1}^N.
\label{t113}
\end{align}
Now we shift the contours from $i\R-\e$ to $i\R+\e_A$ where
$a_j<\e_A<a_j+1$~$j=1,\cdots,N$. Noting 
\begin{align}
&\frac{1}{2\pi i}
\int_{i\R-\e}
\frac{ds q^{a_j}}{(q;q)_{\infty}}
	\frac{(-\zeta)^{s}\Pi(q^s;q^\a,t)}{(-\zeta)^{a_j}\Pi(q^{a_j};q^{\a},t)}
\cdot
\frac{\prod_{m=1}^N(q^{s-a_m};q)_{\infty}}{\prod_{\substack{m=1\\ m\neq j}}^N
(q^{a_j-a_m};q)_{\infty}}
\frac{q^{a_i}}
{q^s-q^{a_i}}
\frac{\pi}{\sin\pi(s-a_j)}
\notag
\\
&=\delta_{i,j}
+\frac{1}{2\pi i}
\int_{i\R+\e_A}
\frac{ds q^{a_j}}{(q;q)_{\infty}}
	\frac{(-\zeta)^{s}\Pi(q^s;q^\a,t)}{(-\zeta)^{a_j}\Pi(q^{a_j};q^\a,t)}
\cdot
\frac{\prod_{m=1}^N(q^{s-a_m};q)_{\infty}}{\prod_{\substack{m=1\\ m\neq j}}^N
(q^{a_j-a_m};q)_{\infty}}
\frac{q^{a_i}}
{q^s-q^{a_i}}
\frac{\pi}{\sin\pi(s-a_j)},
\label{t114}
\end{align}
and using the relation,
\begin{align}
\frac
{\pi (-\zeta)^y}
{\sin\pi y}
=
\int_{-\infty}^{\infty}
dx
\frac
{-\zeta e^{xy}}
{-\zeta+e^x},
\label{t115}
\end{align}
for $0<y<1$, we have
\begin{align}
\left\langle\frac{1}{(\zeta q^{\l^{(N)}_N};q)_{\infty}}\right\rangle
=
\det\left(
\delta_{i,j}
+
\int_{-\infty}^{\infty}dx A(i,x)B(x,j)
\right)_{i,j=1}^N,
\label{t116}
\end{align}
where
\begin{align}
&
A(i,x)=
\int_{i\R+\e_A}
\frac{ds}{2\pi i}
\frac{e^{sx-\delta |s|}\Pi(q^s;q^\a,t)q^{a_i}}{q^s-q^{a_i}}
\prod_{m=1}^N (q^{s-a_m};q)_{\infty},
\label{t117}
\\
&
B(x,i)=
\frac{e^{-a_j x}}{(q;q)_{\infty}\Pi(q^{a_j};q^\a,t)}
\prod_{m=1}^N \frac{1}{(q^{s-a_m};q)_{\infty}}
\cdot
\frac{-\zeta}{-\zeta+e^x}.
\label{t118}
\end{align}
Here in~\eqref{t117}, we introduced the the convergence factor $\delta=0+$.
Noting the basic property of the determinant, we see that this 
is equal to
\begin{align}
\det\left(
\delta_{i,j}
+
\int_{-\infty}^{\infty}dx A(i,x)B(x,j)
\right)_{i,j=1}^N
=
\det\left(1+BA\right)_{L^2(\R)},
\label{t119}
\end{align}
where RHS is the Fredholm determinant on $L^2(\R)$
with the kernel
\begin{align}
&(BA)(x_1,x_2)
\notag
\\
&=
\frac{-\zeta}{-\zeta+e^{x_1}}
\sum_{j=1}^N
\int_{i\R+\e_A}
\frac{ds}{2\pi i}
\frac{e^{(q^s-q^{a_j})t+sx_2-a_jx_1-\delta |s|}}{(q;q)_{\infty}}
\frac{(q^{\a_m-a_j};q)_{\infty}}{(q^{\a_m-s};q)_{\infty}}
\cdot
\frac
{\prod_{m=1}^N(q^{s-a_m};q)_{\infty}}
{\prod_{\substack{m=1\\m\neq j}}(q^{a_j-a_m};q)_{\infty}}
\cdot
\frac{q^{a_j}}{q^s-q^{a_j}}.
\end{align}
Rewriting the summation over $j=1,\cdots,N$ as the contour integral
enclosing $a_j,~j=1,\cdots,N$ positively (this contour is denoted by $D$),
we get
\begin{align}
&~~
\frac{\zeta\log q}{-\zeta+e^x}
\int_D
\frac{dv}{2\pi i}
\int_{i\R+\e_A}
\frac{ds}{2\pi i}
e^{(q^s-q^v)t+sx_2-vx_1-\delta |s|}
\prod_{m=1}^N
\frac
{(q^{\a_m-v};q)_{\infty}(q^{s-a_m};q)_{\infty}}
{(q^{\a_m-s};q)_{\infty}(q^{v-a_m};q)_{\infty}}
\cdot
\frac{q^{v}}{q^s-q^{v}}
\notag
\\
&=
\frac{\zeta\log q}{-\zeta+e^x}
\int_D
\frac{dv}{2\pi i}
\int_{i\R+\e_A}
\frac{ds}{2\pi i}
e^{(q^s-q^v)t+sx_2-vx_1-\delta|s|}
q^{N(s-v)+v}
\prod_{m=1}^N
\frac
{(q^{\a_m-v+1};q)_{\infty}(q^{s-a_m+1};q)_{\infty}}
{(q^{\a_m-s+1};q)_{\infty}(q^{v-a_m+1};q)_{\infty}}
\notag
\\
&
\hspace{7cm}
\times
\left(
\prod_{m=1}^N
\frac
{q^v-q^{\a_m}}{q^s-q^{a_m}}
\frac
{q^s-q^{\a_m}}{q^v-q^{a_m}}
-1
\right)
\frac{1}{q^s-q^{v}}.
\label{t1110}
\end{align}
Here in the equality, we insert ``-1'' in the parenthesis which has no contribution to the integral. 
Substituting the relation,
\begin{align}
\left(
\frac
{q^v-q^{\a_m}}{q^s-q^{a_m}}
\frac
{q^s-q^{\a_m}}{q^v-q^{a_m}}
-1
\right)
\frac{1}{q^s-q^{v}}
=
\sum_{k=0}^{N-1}
\frac
{q^{a_{k+1}}-q^{\a_{k+1}}}
{(q^s-q^{\a_{k+1}})(q^v-q^{a_{k+1}})}
\prod_{\ell=1}^k
\frac
{(q^s-q^{a_\ell})(q^v-q^{\a_\ell})}
{(q^s-q^{\a_\ell})(q^v-q^{a_\ell})}
\label{t1111}
\end{align}
into~\eqref{t1110}, we find $(BA)(x_1,x_2)=-f(x_1)K(x_1,x_2)$
where $f(x)$ and $K(x_1,x_2)$ are defined by~\eqref{t1101} and~\eqref{t1102}
respectively. From this and~\eqref{t116},~\eqref{t119}, we obtain
our desired expression~\eqref{t110}.
\qed

\vspace{3mm}
Our representation~\eqref{t110} can be rewritten in a closer form as in Theorem 3.18 of ~\cite{BC2014}. 
\begin{corollary}
\begin{align}
&\left\langle\frac{1}{(\zeta q^{\l_N};q)_{\infty}}\right\rangle
=
\det\left(
1
+
K_{\zeta}
\right)_{L^2(C_a)},
\label{c120}
\end{align}
where
\begin{align}
K_{\zeta}(v_1,v_2)
=
\frac{-1}{2\pi i}
\int_{i\R+\e} ds
\frac{(-\zeta)^s\pi}{\sin\pi s}
\frac
{e^{(q^{v_1+s}-q^{v_1})t}}
{q^{s+v_1}-q^{v_2}}
\prod_{m=1}^N
\frac
{(q^{s+v_1-a_m};q)_{\infty}(q^{\a_m-v_1};q)_{\infty}}
{(q^{v_1-a_m};q)_{\infty}(q^{\a_m-s-v_1};q)_{\infty}}.
\label{c121d}
\end{align}
\end{corollary}

\noindent
{\bf Remark.} With $\a_1=\a$ and $\a_m\rightarrow\i,~m=2,3,\cdots$,
the factor $\prod_{m=2}^N (q^{\a_m-v_1};q)_\i/(q^{\a_m-x};q)_\i$ becomes unity
and~\eqref{c121} is reduced to the kernel of our Theorem 1 
under $q^{\a}\to\a$ and the change of variables $q^{v_i}=w_i,i=1,2$.

\smallskip
\noindent
{\bf Proof.}
Getting back to LHS of~\eqref{t1110}, we write it as 
$\int_D\frac{dv}{2\pi i} C(x_2,v)D(v,x_1)$ where
\begin{align}
&C(x,v)=
\frac{\log q}{2\pi i}
\int_{i\R+\e_A}ds
\frac{e^{s x+q^s t-\delta |s|}}{q^s-q^{v}}
\prod_{m=1}^N
\frac
{(q^{s-a_m};q)_{\infty}}
{(q^{\a_m-s};q)_{\infty}},
\label{c1202}
\\
&
D(v,x)=
\frac{\zeta}{-\zeta+e^x}
e^{-v x-q^v t}
\prod_{m=1}^N
\frac
{(q^{\a_m-v};q)_{\infty}}
{(q^{v-a_m};q)_{\infty}}.
\label{c1203}
\end{align}
From these equations with~\eqref{t116} and~\eqref{t119}, we have
\begin{align}
\left\langle\frac{1}{(\zeta q^{\l_N};q)_{\infty}}\right\rangle
=
\det(1+ CD)_{L^2(\R)}=\det(1+DC)_{L^2(C_a)}.
\label{c1204}
\end{align}
We easily see that $(DC)(v_1,v_2)=K_{\zeta}(v_1,v_2)$~\eqref{c121d}.
\qed

\appendix
\section{Proof of~\eqref{p105}}
\label{a}
In this appendix, we prove~\eqref{p105} in the case
$(j_1,j_2,\cdots,j_N)=(1,2,\cdots,N)$. The other case can be proved 
in a parallel way. Hereafter we write ${\bf a}=(a_1,\cdots,a_{N_1})$
and $\ba=(\a_{N_1+1},\cdots,\a_N)$

First we will give an explicit expression for
$f({\bf a},\ba, {\bf n},\ell,N_1)$ defined below~\eqref{p1004}.
For this purpose we rewrite~\eqref{p60}
as
\begin{align}
\left\langle\frac{1}{(\zeta q^{\l_N};q)_{\infty}}\right\rangle
&=\frac{-(q;q)_{\infty}^{N-2}}{(2\pi i)^NN!\Pi(q^a;q^\a,t)}
\int_{i\R -\e}ds\int_{\T^{N-1}}\prod_{j=2}^N\frac{dw_j}{w_j}\cdot
\frac{\pi(q^{s-A+1};q)_{\infty}\left(q^{A-s};q\right)_{\infty}}{\sin \pi(s-A)}
\notag
\\
&
\times
{\prod_j}^{(1)}e^{w_jt}(-\zeta)^{\frac{\log w_j}{\log q}}
{\prod_{i,j}}^{(1)}
\frac{1}
{\left(\frac{q^{a_i}}{w_j};q\right)_{\infty}\left(\frac{q^{\a_i}}{w_j};q\right)_{\infty}}
{\prod_{i<j}}^{(1)}\left(\frac{w_i}{w_j};q\right)_{\infty}\left(\frac{w_j}{w_i};q\right)_{\infty}
\notag\\
&
\times
{\prod_j}^{(2)}e^{w_jt}(-\zeta)^{\frac{\log w_j}{\log q}}
{\prod_{i,j}}^{(2)}
\frac{1}
{\left(\frac{q^{a_i}}{w_j};q\right)_{\infty}\left(\frac{q^{\a_i}}{w_j};q\right)_{\infty}}
{\prod_{i<j}}^{(2)}\left(\frac{w_i}{w_j};q\right)_{\infty}\left(\frac{w_j}{w_i};q\right)_{\infty}
\notag\\
&
\times
{\prod_{i,j}}^{(3)}\frac{\left(\frac{w_i}{w_j};q\right)_{\infty}\left(\frac{w_j}{w_i};q\right)_{\infty}}{\left(\frac{q^{a_i}}{w_j};q\right)_{\infty}\left(\frac{q^{a_j}}{w_i};q\right)_{\infty}\left(\frac{q^{\a_i}}{w_j};q\right)_{\infty}\left(\frac{q^{\a_j}}{w_i};q\right)_{\infty}},
\label{a11}
\end{align}
where we used the notations
\begin{align}
&{\prod_j}^{(1)}=\prod_{j=1}^{N_1},
~~{\prod_{i,j}}^{(1)}=\prod_{i,j=1}^{N_1},
~~{\prod_{i<j}}^{(1)}=\prod_{1\le i<j\le N_1},
\notag\\
&{\prod_j}^{(2)}=\prod_{j=N_1+1}^{N},
~~{\prod_{i,j}}^{(2)}=\prod_{i,j=N_1+1}^{N},
~~{\prod_{i<j}}^{(2)}=\prod_{N_1+1\le i<j\le N},
~~{\prod_{i,j}}^{(3)}=\prod_{i=1}^{N_1}\prod_{j=N_1+1}^N.
\label{a12}
\end{align}
From~\eqref{a11}, we find that $f({\bf a},\ba, {\bf n}, \ell,N_1)$
can be written as
\begin{align}
f({\bf a},\ba, {\bf n}, \ell,N_1)=\frac{1}{N!\Pi(q^a;q^\a,t)}P^{(1)}P^{(2)}P^{(3)}.
\end{align}
Here
\begin{align}
&P^{(1)}=
{\prod_j}^{(1)}
\frac{e^{q^{a_j+n_j}t}\zeta^{n_j}}{(q;q)_{n_j}}
\cdot
{\prod_{i,j}}^{(1)}
\frac{1}
{\left(q^{\a_i-a_j-n_j};q\right)_{\infty}}
\cdot
{\prod_{i<j}}^{(1)}
\frac
{\left(q^{a_i-a_j+n_i-n_j};q\right)_{\infty}\left(q^{a_j-a_i+n_j-n_i};q\right)_{\infty}}
{q^{n_in_j}\left(q^{a_i-a_j-n_j};q\right)_{\infty}\left(q^{a_j-a_i-n_i};q\right)_{\infty}},
\label{a14}
\\
&P^{(2)}=\frac{\pi}{\log q}
{\prod_j}^{(2)}
\frac{e^{q^{\a_j+n_j}t}(-1)^{\a_j-a_j}\zeta^{\a_j-a_j+n_j}}{(q;q)_{n_j}}
\cdot
{\prod_{i,j}}^{(2)}
\frac{q^{(a_i-\a_i)n_j}}
{\left(q^{a_i-\a_j-n_j};q\right)_{\infty}}
\notag
\\
&\hspace{5mm}
\cdot
{\prod_{i<j}}^{(2)}
\frac
{\left(q^{\a_i-\a_j+n_i-n_j};q\right)_{\infty}\left(q^{\a_j-\a_i+n_j-n_i};q\right)_{\infty}}
{q^{n_in_j}\left(q^{\a_i-\a_j-n_j};q\right)_{\infty}\left(q^{\a_j-\a_i-n_i};q\right)_{\infty}}
\cdot
\frac
{
(-\zeta)^{\frac{2\pi i\ell}{\log q}}
(q^{\a^{(2)}-A^{(2)}+1};q)_{\infty}
(q^{A^{(2)}-\a^{(2)}};q)_{\infty}
}
{\sin(\a^{(2)}-A^{(2)}+\frac{2\pi i \ell}{\log q})(q;q)_{\infty}^2},
\label{a15}
\\
&P^{(3)}=
{\prod_{i,j}}^{(3)}
\frac
{
q^{(a_j-\a_j)n_i}
(q^{a_i-\a_j+n_i-n_j};q)_{\infty}
(q^{\a_j-a_i+n_j-n_i};q)_{\infty}
}
{
q^{n_in_j}
(q^{a_i-\a_j-n_j};q)_{\infty}
(q^{a_j-a_i-n_i};q)_{\infty}
(q^{\a_i-\a_j-n_j};q)_{\infty}
(q^{\a_j-a_i-n_i};q)_{\infty}
}.
\label{a16}
\end{align}
where 
$\a^{(2)}=\sum_{j=N_1+1}^N\a_j$, $A^{(2)}=\sum_{j=N_1+1}^N a_j$.

Rewriting rhs of~\eqref{p100} in a similar manner,
we have an explicit expression for $g({\bf a},\ba,{\bf n},\bl,N_1)$
defined below~\eqref{p1003},
\begin{align}
g({\bf a},\ba,{\bf n},\bl,N_1)=\frac{1}{N!}Q^{(1)}Q^{(2)}Q^{(3)},
\label{a18}
\end{align}
where
\begin{align}
&Q^{(1)}=
{\prod_j}^{(1)}
\frac{e^{q^{a_j+n_j}t}\zeta^{n_j}}{(q;q)_{n_j}q^{(N_1-1)a_j}}
\cdot
{\prod_{i,j}}^{(1)}
\frac{1}
{\left(q^{\a_i-a_j-n_j};q\right)_{\infty}}
\notag
\\
&
\hspace{1cm}
\times
{\prod_{i<j}}^{(1)}
\frac
{\left(q^{a_i-a_j+n_i+1};q\right)_{\infty}\left(q^{a_j-a_i+n_j+1};q\right)_{\infty}
(q^{a_j+n_j}-q^{a_i+n_i})(q^{a_i}-q^{a_j})
}
{
(q^{a_i-a_j};q)_{\infty}(q^{a_j-a_i};q)_{\infty}
},
\label{a19}
\\
&Q^{(2)}=
{\prod_j}^{(2)}
\frac{
\pi q^{\frac{n_j(n_j+1)}{2}}e^{q^{\a_j+n_j}t}\zeta^{\a_j+n_j+\frac{2\pi i\ell_j}{\log q}-a_j}
}
{
\log q (q;q)_{n_j}(q;q)_{\infty}^2q^{(N-N_1-1)a_j}
}
\cdot
{\prod_{i,j}}^{(2)}
\frac{(q^{\a_i-a_j+n_i+1};q)_{\infty}}
{\sin\pi\left(\a_i-a_j+\frac{2\pi i\ell_i}{\log q}\right)}
\notag
\\
&
\hspace{1cm}
\times
{\prod_{i<j}}^{(2)}
\frac
{
\sin\pi\left(\a_j-\a_i+\frac{2\pi i(\ell_j-\ell_i)}{\log q}\right)
\sin\pi\left(a_j-a_i\right)
(q^{\a_j+n_j}-q^{\a_i+n_i})(q^{a_j}-q^{a_i})
}
{
(q^{\a_i-\a_j-n_j};q)_{\infty}(q^{\a_j-\a_i-n_i};q)_{\infty}
(q^{a_i-a_j};q)_{\infty}(q^{a_j-a_i};q)_{\infty}
},
\label{a110}
\\
&Q^{(3)}=
{\prod_{i,j}}^{(3)}
\frac
{
(q^{a_i-a_j+n_i+1};q)_{\infty}
(q^{\a_j-a_i+n_j+1};q)_{\infty}
(q^{\a_j+n_j}-q^{a_i+n_i})
(q^{a_i}-q^{a_j})
}
{
q^{a_i+a_j}
(q^{\a_i-\a_j-n_j};q)_{\infty}
(q^{\a_j-a_i-n_i};q)_{\infty}
(q^{a_i-a_j};q)_{\infty}
(q^{a_j-a_i};q)_{\infty}
}.
\label{a111}
\end{align}

In the following we will show that
\begin{align}
P^{(1)}=Q^{(1)},
~\sum_{
\substack{
\ell_{N_1+1},\cdots,\ell_N=-\infty
\\
\ell_{N_1+1}+\cdots+\ell_N=\ell
}
}^{\infty}
P^{(2)}=Q^{(2)},
~P^{(3)}=Q^{(3)},
\label{a112}
\end{align}
from which~\eqref{p105} readily follows.
 
For the proof of $P^{(1)}=Q^{(1)}$, it is sufficient to show
\begin{multline}
\frac{
(q^{a_i-a_j+n_i-n_j};q)_{\infty}
(q^{a_j-a_i+n_j-n_i};q)_{\infty}
}
{
q^{n_in_j}
(q^{a_i-a_j-n_j};q)_{\infty}
(q^{a_j-a_i-n_i};q)_{\infty}
}
\\
=
\frac{
(q^{a_i-a_j+n_i+1};q)_{\infty}
(q^{a_j-a_i+n_j+1};q)_{\infty}
(q^{a_j+n_j}-q^{a_i+n_i})
(q^{a_i}-q^{a_j})
}
{
q^{a_i+a_j}
(q^{a_i-a_j};q)_{\infty}
(q^{a_j-a_i};q)_{\infty}
}.
\label{a113}
\end{multline}
We find that lhs of \eqref{a113} becomes
\begin{align}
\frac{
(q^{a_i-a_j+n_i-n_j};q)_{n_j+1}
(q^{a_j-a_i+n_j-n_i};q)_{n_i+1}
}
{
q^{n_in_j}
(q^{a_i-a_j-n_j};q)_{n_j}
(q^{a_j-a_i-n_i};q)_{n_i}
}
\times
\frac{
(q^{a_i-a_j+n_i+1};q)_{\infty}
(q^{a_j-a_i+n_j+1};q)_{\infty}
}
{
(q^{a_i-a_j};q)_{\infty}
(q^{a_j-a_i};q)_{\infty}
}.
\label{a114}
\end{align}
Applying the relation,
\begin{align}
(x;q)_n=(-x)^nq^{\frac{n(n-1)}{2}}\left(x^{-1}q^{1-n};q\right)_n,
\label{a115}
\end{align}
to $(q^{a_j-a_i+n_j-n_i};q)_{n_i+1}$ and $(q^{a_j-a_i-n_i};q)_{n_i}$
in~\eqref{a114},
we see that the first factor is written as
\begin{multline}
\frac{
(q^{a_i-a_j+n_i-n_j};q)_{n_j+1}
(q^{a_i-a_j-n_j};q)_{n_i+1}
(-q^{a_j-a_i+n_j-n_i})^{n_i+1}
q^{\frac{n_i(n_i+1)}{2}}
}
{
(q^{a_i-a_j-n_j};q)_{n_j}
(q^{a_i-a_j+1};q)_{n_i}
(-q^{a_j-a_i-n_i})^{n_i}
q^{\frac{n_i(n_i-1)}{2}+n_in_j}
}
\\
=
\frac{(q^{a_j+n_j}-q^{a_i+n_i})(q^{a_i}-q^{a_j})}{q^{a_i+a_j}},
\label{a116}
\end{multline}
where we used
\begin{align}
(q^{a_i-a_j+n_i-n_j};q)_{n_j+1}
(q^{a_i-a_j-n_j};q)_{n_i}
=(q^{a_i-a_j-n_j};q)_{n_i+n_j}
=
(q^{a_i-a_j-n_j};q)_{n_j+1}(q^{a_i-a_j+1};q)_{n_i}.
\label{a117}
\end{align}
Substituting the above relation to~\eqref{a114}, we arrive 
at~\eqref{a113}.

For proving $P^{(3)}=Q^{(3)}$ it is sufficient to show
\begin{align}
\frac
{
q^{a_i}
(q^{a_i-\a_j+n_i-n_j+1};q)_{\infty}
(q^{\a_j-a_i+n_j-n_i};q)_{\infty}
}
{
q^{(n_i+1)(\a_j+n_j)}
(q^{a_i-\a_j-n_j};q)_{\infty}
(q^{\a_j-a_i+n_j+1};q)_{\infty}
}
=
\frac
{
-(q^{a_i-a_j+n_i+1};q)_{\infty}
(q^{a_j-a_i-n_i};q)_{\infty}
}
{
q^{a_jn_i}
(q^{a_i-a_j+1};q)_{\infty}
(q^{a_j-a_i};q)_{\infty}
}.
\label{a118}
\end{align}
Using~\eqref{a115},
we see that the lhs becomes
\begin{align}
\frac
{
q^{a_i}
(q^{\a_j-a_i+n_j-n_i};q)_{n_i+1}
}
{
q^{(n_i+1)(\a_j+n_j)}
(q^{a_i-\a_j-n_j};q)_{n_i+1}
}
&=
\frac
{
q^{a_i}
(q^{a_i-\a_j-n_j};q)_{n_i+1}(-q^{\a_j-a_i+n_j-n_i})^{n_i+1}
q^{\frac{n_i(n_i+1)}{2}}
}
{
q^{(n_i+1)(\a_j+n_j)}
(q^{a_i-\a_j-n_j};q)_{n_i+1}
}
\notag
\\
&=
(-1)^{n_i+1}q^{-(a_i+n_i)n_i}q^{\frac{n_i(n_i-1)}{2}}.
\label{a119}
\end{align}
Also from~\eqref{a115}, the rhs becomes
\begin{align}
-
\frac
{
(q^{a_j-a_i-n_i};q)_{n_i}
}
{
q^{a_jn_i}
(q^{a_i-a_j+1};q)_{n_i}
}
=
-
\frac
{
(q^{a_i-a_j+1};q)_{n_i}
(-q^{a_j-a_i-n_i})^{n_i}q^{\frac{n_i(n_i-1)}{2}}
}
{
q^{a_jn_i}
(q^{a_i-a_j+1};q)_{n_i}
}
=
(-1)^{n_i+1}q^{-(a_i+n_i)n_i}q^{\frac{n_i(n_i-1)}{2}}.
\label{a120}
\end{align}
In these equations~\eqref{a119} and~\eqref{a120}, we used
\eqref{a115}.

Finally we show $P^{(2)}=Q^{(2)}$.  For the proof, it is sufficient to show
\begin{align}
&~~
\sum_{\substack
{\ell_{1},\cdots,\ell_{N}=-\infty
\\
\ell_{1}+\cdots+\ell_{N}=\ell
}}^{\infty}
\frac{
\prod_{1\le i<j\le N}
\sin\pi \left(z_j-z_i+\frac{2\pi i (\ell_j-\ell_i)}{\log q}\right)
}
{
\prod_{i,j=1}^N
\sin\pi \left(z_i-a_j+\frac{2\pi i  \ell_i}{\log q}\right)
}
\notag\\
&=
\frac{
(
(q;q)_{\infty}^2
q^{A}\log q
)^{N-1}
}
{
\pi^{N-1}
\prod_{1\le i<j\le N}
(q^{z_j}-q^{z_i})(q^{a_j}-q^{a_i})
\sin\pi (a_j-a_i)
}
\frac
{
(q^{Z-A+1};q)_{\infty}
(q^{A-Z};q)_{\infty}
}
{
\sin\pi\left(Z-A+\frac{2\pi i \ell}{\log q}\right)
}
\notag\\
&
\times
\frac
{
\prod_{1\le i<j\le N}
(q^{z_i-z_j};q)_{\infty}
(q^{z_j-z_i};q)_{\infty}
(q^{a_i-a_j};q)_{\infty}
(q^{a_j-a_i};q)_{\infty}
}
{
\prod_{i,j=1}^N
(q^{z_i-a_j+1};q)_{\infty}
(q^{a_i-z_j};q)_{\infty}
},
\label{a121}
\end{align}
where $z_j,~a_j\in \C,~j=1,\cdots,N$, $\ell\in\Z$, $Z=z_1+\cdots+z_N$, and
$A=a_1+\cdots+a_N$.
We show this equality by mathematical induction. For $N=1$,
we find that both side hand becomes
$1/\sin(z_1-a_1+\frac{2\pi i \ell}{\log q})$.
Suppose that it holds for $N-1$. Then for $N$, we will show that
both hand side in~\eqref{a121} have the same Laurant series with
repect to $z_1$. (Note that they are both meromorphic functions with 
respect to $z_1,\cdots,z_N$.) 

Notice that for a meromorphic function $f(z)$ with poles $z=p_j,~j=1,2,\cdots$,
their exists a unique Laurent expansion,
\begin{align}
f(z)=f(0)+\sum_{k}\left[f_k\left(\frac{1}{z-p_k}\right)-f_k\left(\frac{1}{-p_k}\right)
\right],
\label{a122}
\end{align}
where $f_k(1/(z-p_k))$ is the principal part at $z=p_k$. Let us denote
lhs of~\eqref{a121} as $f(z_1,\cdots, z_N)$. It has the poles at
$z_1=a_j+n_1-2\pi \ell_1/\log q$ where $j=1,\cdots,N$ and $n_1,\ell_1\in
\Z$ and is expanded as
\begin{multline}
f(z_1,z_2,\cdots,z_N)-f(0,z_2,\cdots,z_N)
\\
=
\sum_{j=1}^N 
\sum_{n_1=-\infty}^{\infty}
\sum_{\ell_1=-\infty}^{\infty}
\left[
f_{j,n_1,\ell_1}
\left(
\frac{1}{z_1-a_j-n_1+
\frac{2\pi i \ell_1}{\log q}
}
\right)
-
f_{j,n_1,\ell_1}
\left(
\frac{1}{-a_j-n_1+
\frac{2\pi i \ell_1}{\log q}
}
\right)
\right]
\label{a123}.
\end{multline}
We can write the principal parts
$
f_{j,n_1,\ell_1}
\left(
\frac{1}{z_1-a_1-n_1+
\frac{2\pi i \ell_1}{\log q}
}
\right)
$
explicitly. For $j=1$, it is given as 
\begin{multline}
f_{1,n_1,\ell_1}
\left(
\frac{1}{z_1-a_1-n_1+
\frac{2\pi i \ell_1}{\log q}
}
\right)
\\
=
\frac{1}{z_1-a_1-n_1+
\frac{2\pi i \ell_1}{\log q}}
\frac{(-1)^{n_1}}{\prod_{j=2}^N\sin\pi(a_1-a_j)}
\sum_{
\substack
{\ell_2,\cdots,\ell_N=-\infty
\\
\ell_2+\cdots+\ell_N=\ell-\ell_1
}
}^{\infty}
\frac{
\prod_{2\le i<j\le N}
\sin\pi \left(z_j-z_i+\frac{2\pi i (\ell_j-\ell_i)}{\log q}\right)
}
{
\prod_{i,j=2}^N
\sin\pi \left(z_i-a_j+\frac{2\pi i  \ell_i}{\log q}\right)
}.
\label{a124}
\end{multline}

Let $g(z_1,\cdots,z_N)$ be rhs of~\eqref{a121}. It can be written as
\begin{multline}
g(z_1,z_2,\cdots,z_N)-g(0,z_2,\cdots,z_N)
\\
=
\sum_{j=1}^N 
\sum_{n_1=-\infty}^{\infty}
\sum_{\ell_1=-\infty}^{\infty}
\left[
g_{j,n_1,\ell_1}
\left(
\frac{1}{z_1-a_j-n_1+
\frac{2\pi i \ell_1}{\log q}
}
\right)
-
g_{j,n_1,\ell_1}
\left(
\frac{1}{-a_j-n_1+
\frac{2\pi i \ell_1}{\log q}
}
\right)
\right],
\label{a125}
\end{multline}
where the principal part for the case $j=1$ is given as
\begin{align}
&~g_{1,n_1,\ell_1}
\left(
\frac{1}{z_1-a_1-n_1+
\frac{2\pi i \ell_1}{\log q}
}
\right)
=
\frac{1}
{
z_1-a_1-n_1+\frac{2\pi i \ell_1}{\log q}
}
\frac{(-1)^{n_1}}{\prod_{j=2}^N\sin\pi(a_1-a_j)}
\notag
\\
&\times
\frac{
(
\log q(q;q)_{\infty}^2
q^{A^{(N-1)}}
)^{N-2}
}
{
\pi^{N-2}
\prod_{2\le i<j\le N}
(q^{z_j}-q^{z_i})(q^{a_j}-q^{a_i})
\sin\pi (a_j-a_i)
}
\frac
{
(q^{Z^{(N-1)}-A^{(N-1)}+n_1+1};q)_{\infty}
(q^{A^{(N-1)}-Z^{(N-1)}-n_1};q)_{\infty}
}
{
\sin\pi\left(Z^{(N-1)}-A^{(N-1)}+\frac{2\pi i \ell}{\log q}\right)
}
\notag\\
&
\times
\frac
{
\prod_{2\le i<j\le N}
(q^{z_i-z_j};q)_{\infty}
(q^{z_j-z_i};q)_{\infty}
(q^{a_i-a_j};q)_{\infty}
(q^{a_j-a_i};q)_{\infty}
}
{
\prod_{i,j=2}^N
(q^{z_i-a_j+1};q)_{\infty}
(q^{a_i-z_j};q)_{\infty}
}
\notag
\\
&\times
\frac
{1}
{(q^{n_1+1};q)_{\infty}}
\prod_{j=2}^N
\frac{
(q^{a_1+n_1-z_j};q)_{\infty}
(q^{z_j-a_1-n_1};q)_{\infty}
(q^{a_1-a_j};q)_{\infty}
(q^{a_j-a_1};q)_{\infty}
}
{
(q^{a_1+n_1-a_j+1};q)_{\infty}
(q^{z_j-a_1+1};q)_{\infty}
(q^{a_1-z_j};q)_{\infty}
(q^{a_j-a_1-n_1};q)_{\infty}
}
\notag
\\
&\times
\frac
{\log q (q;q)_{\infty}^2q^{a_1(N-1)+A^{(N-1)}}
}
{
(-1)^{N-1}
\prod_{j=2}^N (q^{a_j}-q^{a_1})
(q^{z_j}-q^{a_1+n_1})
}
\frac
{
(-1)^{n_1}
q^{\frac{n_1(n_1+1)}{2}}
}
{
 (q;q)_{n_1}(q;q)_{\infty}\log q
},
\label{a126}
\end{align}
where $Z^{(N-1)}=z_2+\cdots+z_N,$
$A^{(N-1)}=a_2+\cdots+a_N$.
Noting the following deformation,
\begin{align}
&~~
\prod_{j=2}^N
\frac{
(q^{a_1+n_1-z_j};q)_{\infty}
(q^{z_j-a_1-n_1};q)_{\infty}
(q^{a_1-a_j};q)_{\infty}
(q^{a_j-a_1};q)_{\infty}
}
{
(q^{a_1+n_1-a_j+1};q)_{\infty}
(q^{z_j-a_1+1};q)_{\infty}
(q^{a_1-z_j};q)_{\infty}
(q^{a_j-a_1-n_1};q)_{\infty}
}
\notag
\\
&=
\prod_{j=2}^N
\frac{
(q^{z_j-a_1-n_1};q)_{n_1+1}
(q^{a_1-a_j};q)_{n_1+1}
}
{
(q^{a_1-z_j};q)_{n_1}
(q^{a_j-a_1-n_1};q)_{n_1}
}
\notag
\\
&=
\prod_{j=2}^N
\frac{
(-q^{z_j-a_1-n_1})^{n_1+1}
q^{\frac{n_1(n_1+1)}{2}}
(q^{a_1-z_j};q)_{n_1+1}
(q^{a_1-a_j};q)_{n_1+1}
}
{
(-q^{a_j-a_1-n_1})^{n_1}
q^{\frac{n_1(n_1-1)}{2}}
(q^{a_1-z_j};q)_{n_1}
(q^{a_1-a_j+1};q)_{n_1}
}
\notag
\\
&=
\frac{(-1)^{N-1}q^{(Z^{(N-1)}-A^{(N-1)})n_1}}
{
q^{(N-1)a_1+A^{(N-1)}}
}
\prod_{j=2}^N
(q^{a_j}-q^{a_1})
(q^{z_j}-q^{a_1+n_1}),
\label{a127}
\end{align}
where in the second equality we used the relation~\eqref{a115}, we arrive at
\begin{align}
&~g_{1,n_1,\ell_1}
\left(
\frac{1}{z_1-a_j-n_1+
\frac{2\pi i \ell_1}{\log q}
}
\right)
=
\frac{1}
{
z_1-a_1-n_1+\frac{2\pi i \ell_1}{\log q}
}
\frac{(-1)^{n_1}}{\prod_{j=2}^N\sin\pi(a_1-a_j)}
\notag
\\
&\times
\frac{
(
\log q(q;q)_{\infty}^2
q^{A^{(N-1)}}
)^{N-2}
}
{
\pi^{N-2}
\prod_{2\le i<j\le N}
\sin\pi (a_j-a_i)(q^{z_j}-q^{z_i})(q^{a_j}-q^{a_i})
}
\frac
{
(q^{Z^{(N-1)}-A^{(N-1)}+1};q)_{\infty}
(q^{A^{(N-1)}-Z^{(N-1)}};q)_{\infty}
}
{
\sin\pi\left(Z^{(N-1)}-A^{(N-1)}+\frac{2\pi i \ell}{\log q}\right)
}
\notag\\
&
\times
\frac
{
\prod_{2\le i<j\le N}
(q^{z_i-z_j};q)_{\infty}
(q^{z_j-z_i};q)_{\infty}
(q^{a_i-a_j};q)_{\infty}
(q^{a_j-a_i};q)_{\infty}
}
{
\prod_{i,j=2}^N
(q^{z_i-a_j+1};q)_{\infty}
(q^{a_i-z_j};q)_{\infty}
}.
\label{a129}
\end{align}
By the assumption of the mathematical induction for the case $N-1$, 
we can check the equivalence between~\eqref{a124} and~\eqref{a129}
and the case of other $j$'s can be shown in a same way.
Thus the problem to show the equivalence between~\eqref{a123} and~\eqref{a125}
reduces to the one to show $f(0,z_2,z_3,\cdots,z_N)=g(0,z_2,z_3,\cdots,z_N)$.
Applying the same strategy to this relation, the problem further reduces to
$f(0,0,z_3,\cdots,z_N)=g(0,0,z_3,\cdots,z_N)$. However we easily find that
\begin{align}
f(0,0,z_3,\cdots,z_N)
=
g(0,0,z_3,\cdots,z_N)
=0,
\label{a128}
\end{align}
due to the antisymmetry. This completes the proof of~\eqref{a121}.

\smallskip
\qed

\smallskip
\noindent
{\bf Acknowledgements.}
The work of T.I. and T.S. is supported by JSPS KAKENHI Grant Numbers JP25800215, JP16K05192 and 
JP25103004, JP14510499, JP15K05203, JP16H06338 respectively. 

\bibliographystyle{amsplain}
\providecommand{\bysame}{\leavevmode\hbox to3em{\hrulefill}\thinspace}
\providecommand{\MR}{\relax\ifhmode\unskip\space\fi MR }
\providecommand{\MRhref}[2]{%
  \href{http://www.ams.org/mathscinet-getitem?mr=#1}{#2}
}
\providecommand{\href}[2]{#2}

\end{document}